\documentclass[reprint,aps,prd,onecolumn,notitlepage,showpacs,nofootinbib,preprintnumbers,superscriptaddress]{revtex4-1}
\usepackage{amsmath}
\usepackage{amsfonts}
\usepackage{amssymb}
\usepackage{latexsym}
\usepackage{color}
\usepackage{graphicx}
\usepackage{bm}
\usepackage{epsfig}
\usepackage[english]{babel}
\usepackage{hyperref}
\usepackage{times}
\usepackage{comment}

\def\con{\mathrm{constant}}

\begin{document}

\title{Membrane paradigm of the static black bottle}
\author{Li Li}
\author{Towe Wang}
\email[Electronic address: ]{twang@phy.ecnu.edu.cn}
\affiliation{Department of Physics, East China Normal University,\\
Shanghai 200241, China\\ \vspace{0.2cm}}
\date{\today\\ \vspace{1cm}}
\begin{abstract}
In the membrane paradigm of black holes, it is usually assumed that the normal vector of the stretched horizon has a vanishing acceleration. This assumption breaks down for black bottles, a class of solutions discovered recently in the asymptotically anti-de Sitter spacetime. In this paper, the membrane paradigm is generalized to the stretched horizon with a nonvanishing acceleration of normal vector, and then it is applied to the static black bottle. In this example, the membrane stress tensor and the fluid quantities are similar to those of black holes, while the fluid continuity equation and the Navier-Stokes equation are well satisfied in the near-horizon limit.
\end{abstract}


\maketitle




\section{Introduction}\label{sect-intro}
In the past half century, the black hole thermodynamics has greatly deepened our understanding of gravity theory. However, the dynamics of gravity is not merely thermodynamics. The membrane paradigm of black holes has opened for us another window to gravity---the hydrodynamics. On the event horizon of a black hole, the gravitational equations resemble the low-dimensional fluid continuity equation and the Navier-Stokes equation. The fluid quantities, such as density, pressure, shear and expansion, are encoded on the stretched horizon---a timelike hypersurface slightly outside the event horizon. Algebraically, the black hole membrane paradigm can be derived from Einstein equations \cite{Damour:1978cg,Damour79,Damour82,Price:1986yy,Thorne86}, or alternatively from an action with a surface term \cite{Parikh:1997ma}. The action-based method is convenient for studying more complicated examples. Although the membrane paradigm is normally applied to black holes in the Einstein gravity, it is straightforward to extend it to other gravity theories \cite{Chatterjee:2010gp,Jacobson:2011dz,Kolekar:2011gg,Fischler:2015kro,Zhao:2015inu} and to the Friedmann-Lema\^{i}tre-Robertson-Walker spacetime \cite{Wang:2014aty}. Remarkably, turning to black branes in the asymptotically anti-de Sitter (AdS) spacetime, one can make use of the membrane paradigm and the AdS/CFT correspondence to calculate hydrodynamic quantities of the dual field theory, see \cite{Kovtun:2003wp,Starinets:2008fb,Iqbal:2008by} and references therein.

Topologically, a black hole has a spherical horizon, while a black brane has a planar horizon. In AdS spacetime, there are black hole-like solutions with various nontrivial topologies. In reference \cite{Chen:2016rjt}, Chen and Teo discovered a new class of solutions with bottle-shaped horizons, which they named as black bottle. Contingent on the value of a rotation parameter, the black bottle can be either static or rotating. In the present paper, we will focus on the static black bottle, whose horizon interpolates between a spherical geometry and a hyperbolic one. We intend to build the membrane paradigm of the static black bottle. Unlike black holes or black branes, on the stretched horizon, the black bottle has a nonvanishing acceleration of normal vector, violating the assumption of zero acceleration made in the literature. We will generalize some results in the literature to take this point into consideration.

The paper is organized as follows. After a concise technical review of membrane paradigm in section \ref{sect-rev}, we will reformulate the static solution in spherical coordinates in section \ref{sect-spher}. Then the membrane paradigm will be established in section \ref{sect-memBB} for the static black bottle explicitly. We make a $2+1+1$ split of the spacetime in subsection \ref{subsect-split} by specifying the timelike generator and the spacelike normal of the stretched horizon. Most of the fluid quantities on the membrane will be worked out in subsection \ref{subsect-quant} in spherical coordinates. In subsection \ref{subsect-EF}, the Eddington-Finkelstein (EF) coordinates will be introduced to evaluate the surface gravity and inspect the limit that the stretched horizon tends to the true horizon. We will check the fluid equations in subsection \ref{subsect-flueq}. The main results in section \ref{sect-memBB} will be translated to the Pleba\'nski-Demia\'nski-like coordinates in section \ref{sect-PD}. The paper concludes in section \ref{sect-con} with a summary and discussion of the results. In appendix \ref{app-teh}, we will show that the membrane stress tensor in the Einstein gravity receives no correction from the nonzero acceleration of normal vector.

\section{Review of membrane paradigm}\label{sect-rev}
In this section, we will briefly review some geometric and algebraic details of the membrane paradigm of black holes in the Einstein gravity.

The black hole event horizon $\mathcal{H}$ is a 3-dimensional null hypersurface with a null geodesic generator $l^{a}$. At the event horizon, the geodesic equation is \begin{equation}\label{surfg}
l^{b}\nabla_{b}l^{a}=g_{\mathcal{H}}l^{a}£¬
\end{equation}
where $\nabla_{b}$ is the covariant derivative with respect to metric $g_{ab}$ of the 4-dimensional spacetime. In a stationary spacetime, $l^{a}$ can be taken as the null limit of a timelike Killing vector and then $g_{\mathcal{H}}$ will be the surface gravity at $\mathcal{H}$.

The membrane paradigm is constructed on a timelike membrane $\mathcal{S}$, which is outside $\mathcal{H}$ and very close to it. This membrane is dubbed the stretched horizon. The timelike generator $u^{a}$ and the spacelike normal $n^{a}$ of $\mathcal{S}$ are normalized to unity,
\begin{equation}\label{norm}
n_{a}u^{a}=0,~~~~n^{a}n_{a}=1,~~~~u^{a}u_{a}=-1.
\end{equation}
The metric of $\mathcal{S}$ is given by
\begin{equation}
h_{ab}=g_{ab}-n_{a}n_{b}.
\end{equation}
Then we can define the extrinsic curvature of stretched horizon as
\begin{equation}\label{3k}
K_{ab}=h_{b}^{~d}\nabla_{d}n_{a},
\end{equation}
and the extrinsic curvature scalar $K=g^{ab}K_{ab}$. A 2-dimensional spacelike cross section of $\mathcal{S}$ normal to $u^{a}$ has the metric
\begin{equation}
\gamma_{ab}=h_{ab}+u_{a}u_{b}.
\end{equation}

In the membrane paradigm, we introduce a parameter $\alpha$. When $\alpha\rightarrow 0$, the stretched horizon $\mathcal{S}$ will tend to the event horizon $\mathcal{H}$, and \begin{equation}\label{nulim}
\alpha u^{a}\rightarrow l^{a},~~~~\alpha n^{a}\rightarrow l^{a}.
\end{equation}
At the same time, the aforementioned 2-section of $\mathcal{S}$ will approach a 2-dimensional spacelike cross section of $\mathcal{H}$, and
\begin{equation}\label{Klim}
K_{ab}\rightarrow\alpha^{-1}k_{ab}-\alpha^{-1}g_{\mathcal{H}}u_{a}u_{b}.
\end{equation}
Here $k_{AB}$ is the extrinsic curvature of the 2-dimensional spacelike section of $\mathcal{H}$,
\begin{equation}\label{2k}
k_{AB}=\gamma^{a}_{~A}\gamma^{b}_{~B}\nabla_{b}l_{a}=\frac{1}{2}\mathcal{L}_{l}\gamma_{AB}.
\end{equation}
The notation $\mathcal{L}_{l}$ is the Lie derivative along $l^{a}$. It proves convenient to decompose $k_{AB}$ into a traceless part and a trace,
\begin{equation}\label{2kdec}
k_{AB}=\sigma_{AB}+\frac{1}{2}\theta\gamma_{AB},
\end{equation}
where $\sigma_{AB}$ is the shear of $l_{a}$ and $\theta$ is the expansion.

In reference \cite{Parikh:1997ma}, assuming $n^{a}\nabla_{a}n_{b}=0$, it was proven that the membrane stress tensor
\begin{equation}\label{teh}
t_{ab}=\frac{1}{8\pi G} (Kh_{ab}-K_{ab})
\end{equation}
on $\mathcal{S}$ in the Einstein gravity. As is demonstrated in appendix \ref{app-teh}, this result remains correct even if $n^{a}\nabla_{a}n_{b}\neq0$. The above stress tensor can be decomposed to a form like a viscous fluid,
\begin{equation}\label{tdec}
t_{ab}=\frac{1}{\alpha}\rho u_{a}u_{b}+\frac{1}{\alpha}\gamma_{aA}\gamma_{bB}(p\gamma^{AB}-2\eta\sigma^{AB}-\zeta\theta\gamma^{AB})+\pi^{A}(\gamma_{aA}u_{b}+\gamma_{bA}u_{a}).
\end{equation}
Here we have inserted the renormalization parameter $\alpha$, hence $\rho$, $p$, $\sigma^{AB}$, $\theta$ and $\pi^{A}$ correspond to fluid quantities on $\mathcal{H}$.

Taking the limit $\alpha\rightarrow 0$, we expect the fluid quantities satisfy the $(2+1)$-dimensional fluid continuity equation and the Navier-Stokes equation,
\begin{eqnarray}
\mathcal{L}_{l}\rho+\theta\rho&=&-p\theta+2\eta\sigma_{AB}\sigma^{AB}+\zeta\theta^{2}+T_{ab}l^{a}l^{b},\label{Ray}\\
\gamma_{A}^{~e}\mathcal{L}_{l}\pi_{e}+\pi_{A}\theta&=&-p_{||A}+2(\eta\sigma^{B}_{~A})_{||B}+(\zeta\theta)_{||A}-T_{ab}l^{b}\gamma^{a}_{~A},\label{NS}
\end{eqnarray}
where $T_{ab}$ is the bulk stress tensor, and $||A$ is the 2-covariant derivative with respect to the metric $\gamma_{AB}$.

Some fluid quantities and relations can be known without going into a specific spacetime. Substituting equations \eqref{Klim}, \eqref{2kdec} into equation \eqref{teh}, one gets
\begin{equation}
t_{ab}=-\frac{1}{8\pi G\alpha}\theta u_{a}u_{b}-\frac{1}{8\pi G\alpha}\sigma_{ab}+\frac{1}{8\pi G\alpha}\left(\frac{1}{2}\theta+g_{\mathcal{H}}\right)\gamma_{ab}.
\end{equation}
Identifying it with equation \eqref{tdec}, we can read off
\begin{eqnarray}\label{fluqn1}
\nonumber&&\rho=-\frac{\theta}{8\pi G},~~~~p=\frac{g_{\mathcal{H}}}{8\pi G},~~~~\pi^{A}=0,\\
&&\eta=\frac{1}{16\pi G},~~~~\zeta=-\frac{1}{16\pi G}.
\end{eqnarray}
In subsection \ref{subsect-quant}, we will obtain the other fluid quantities from the details of a static black bottle.

\section{Black bottle in spherical coordinates}\label{sect-spher}
In reference \cite{Chen:2016rjt}, the term ``black bottle'' was coined to refer to a new class of solutions with bottle-shaped event horizons, among which the static solution has the form
\begin{eqnarray}\label{metricxy}
\nonumber ds^2&=&\frac{\ell^2(1-b)}{(x-y)^2}\left[Q(y)dt^2-\frac{dy^2}{Q(y)}+\frac{dx^2}{P(x)}+P(x)d\phi^2\right],\\
\nonumber P(x)&=&1+x-x^2-x^3,\\
Q(y)&=&b+y-y^2-y^3.
\end{eqnarray}
Here $\ell$ is related to the cosmological constant by $\ell^2=-3/\Lambda$. The ranges of parameters and coordinates are
\begin{equation}
b<-5/27,~~~~y<x,~~~~-1<x\leq+1,~~~~0\leq\phi<\pi.
\end{equation}

For our purpose in this paper, it would be helpful to rewrite the above metric in the system of spherical coordinates. This can be achieved by the following transformation of coordinates
\begin{equation}\label{cotrans}
t=\frac{\tau}{\ell\sqrt{2(1-b)}},~~~~x-y=\frac{\ell\sqrt{1-b}}{\sqrt{2}r},~~~~x=\cos\vartheta,~~~~\phi=\frac{\varphi}{2}.
\end{equation}
For simplicity, we will use the abbreviated notation $L=\ell\sqrt{1-b}/\sqrt{2}$. Then it is straightforward to write down the line element of a static black bottle
\begin{eqnarray}\label{metrictr}
\nonumber ds^2&=&-f(r,\vartheta)d\tau^2+\frac{1}{f(r,\vartheta)}\left(dr-\frac{r^2\sin\vartheta}{L}d\vartheta\right)^2+\frac{2r^2}{1+\cos\vartheta}d\vartheta^2+\frac{r^2\sin^2\vartheta}{2}(1+\cos\vartheta)d\varphi^2,\\
f(r,\vartheta)&=&-\frac{L}{2r}+\frac{1}{2}(1+3\cos\vartheta)+\frac{r}{2L}(1-2\cos\vartheta-3\cos^2\vartheta)-\frac{r^2}{2L^2}(b+\cos\vartheta-\cos^2\vartheta-\cos^3\vartheta)
\end{eqnarray}
as well as the ranges of parameters and coordinates
\begin{equation}
b<-5/27,~~~~r>0,~~~~0\leq\vartheta<\pi,~~~~0\leq\varphi<2\pi.
\end{equation}

It is remarkable that the function $f(r,\vartheta)$ can also be written as
\begin{eqnarray}
\nonumber f(r,\vartheta)&=&-\frac{r^2}{2L^2}Q(y),\\
Q(y)&=&b+\left(\cos\vartheta-\frac{L}{r}\right)-\left(\cos\vartheta-\frac{L}{r}\right)^2-\left(\cos\vartheta-\frac{L}{r}\right)^3.
\end{eqnarray}
In the case $b<-5/27$, $Q(y)$ has only one real root $y_1$. As demonstrated in reference \cite{Chen:2016rjt}, this root corresponds to a Killing horizon in the spacetime. Translated into spherical coordinates, the horizon is located at the hypersurface
\begin{equation}\label{th}
\cos\vartheta-\frac{L}{r}=y_1.
\end{equation}
Substituting this equation into metric \eqref{metrictr}, one can plot the same image of black bottle as figure 4 in reference \cite{Chen:2016rjt}. Notice that the shape of black bottle is controlled by  parameter $y_1$ or alternatively $b$, while its area is determined by $y_1$ and $L$ together,
\begin{equation}
A_{BB}=\frac{4\pi L^2}{y_1^2-1}.
\end{equation}
It is easy to check that the above value is consistent with equation (27) in reference \cite{Chen:2016rjt}.

It is interesting to observe that $\partial_{r}Q=(L/r^2)\partial_{y}Q$, $\partial_{\vartheta}Q=-\sin\vartheta\partial_{y}Q$. They can be combined to yield a useful relation
\begin{equation}\label{onf}
\partial_{\vartheta}f=\frac{r^2\sin\vartheta}{L}\left(\frac{2}{r}f-\partial_{r}f\right).
\end{equation}
In this paper, we will not use the explicit expression of $f(r,\vartheta)$. Instead, we will make use of equation \eqref{onf} implicitly.

\section{Membrane paradigm of black bottle}\label{sect-memBB}
\subsection{The $2+1+1$ split}\label{subsect-split}
In this section, we will apply the general setup of section \ref{sect-rev} to a static black bottle. As we will see, although the black bottle has a horizon of unusual topology, its membrane paradigm can be explicitly built as usual. To this end, the first and key step is making a $2+1+1$ split of spacetime, which can be done most conveniently in spherical coordinates.

In the spirit of section \ref{sect-rev}, we wish to rearrange the line element of a black bottle as
\begin{equation}
g_{ab}dx^{a}dx^{b}=-u_{a}u_{b}dx^{a}dx^{b}+n_{a}n_{b}dx^{a}dx^{b}+\gamma_{ab}dx^{a}dx^{b},
\end{equation}
in which $u^{a}$ is proportional to a timelike Killing vector, and $n^{a}$ is normal to the stretched horizon. The most efficient way to do this is comparing the above line element with equation \eqref{metrictr}, which suggests that
\begin{eqnarray}\label{nugamma}
\nonumber&&n_{a}dx^{a}=f^{-1/2}\left(dr-\frac{r^2\sin\vartheta}{L}d\vartheta\right),~~~~u_{a}dx^{a}=-f^{1/2}d\tau,\\
&&\gamma_{AB}dx^{A}dx^{B}=\frac{2r^2}{1+\cos\vartheta}d\vartheta^2+\frac{r^2\sin^2\vartheta}{2}(1+\cos\vartheta)d\varphi^2.
\end{eqnarray}
Accordingly, the metric of stretched horizon is
\begin{equation}\label{h}
h_{ab}dx^{a}dx^{b}=-f(r,\vartheta)d\tau^2+\frac{2r^2}{1+\cos\vartheta}d\vartheta^2+\frac{r^2\sin^2\vartheta}{2}(1+\cos\vartheta)d\varphi^2.
\end{equation}
The one-forms $n_{a}dx^{a}$, $u_{a}dx^{a}$ are dual to vectors
\begin{equation}\label{nu}
n^{a}\partial_{a}=f^{1/2}\partial_{r},~~~~u^{a}\partial_{a}=f^{-1/2}\partial_{\tau},
\end{equation}
meeting the orthonormal condition \eqref{norm}.

We can get a clearer picture of the membrane paradigm from the explicit results above. In the membrane paradigm, the stretched horizon, located slightly outside the event horizon, plays the role of a membrane. The low-dimensional fluid lives on this membrane. From equation \eqref{nugamma}, we notice that $n^{a}$ is the normal vector of hypersurface $\cos\vartheta-L/r=\con$, including the event horizon \eqref{th}. This implies that the stretched horizon of a static black bottle is
\begin{equation}\label{sh}
\cos\vartheta-\frac{L}{r}=y_1+\epsilon,
\end{equation}
where $\epsilon$ is a small positive constant.

For a static black bottle, the null generator of event horizon can be chosen as the timelike Killing vector
\begin{equation}\label{l}
l^{a}\partial_{a}=\partial_{\tau}
\end{equation}
so that the non-affine coefficient $g_{\mathcal{H}}$ in formula \eqref{surfg} is the surface gravity. In practice, this formula can be implemented in the EF coordinates, but not in the spherical coordinates. We will return to this formula in subsection \ref{subsect-EF} and confirm that vector \eqref{l} is a generator of the event horizon. In addition, the limit \eqref{nulim} can be better understood in the EF coordinates. Nevertheless, from equations \eqref{nu} and \eqref{l}, we can guess the renormalization parameter
\begin{equation}\label{alpha}
\alpha=f^{1/2}.
\end{equation}

It is important to point out that the acceleration $a^{b}$ of the normal vector is not zero,
\begin{equation}\label{acc}
a_{b}=n^{a}\nabla_{a}n_{b}=-\frac{r\sin\vartheta}{L}(d\vartheta)_{b},
\end{equation}
violating the assumption $a^{b}=0$ made in the literature. We take this point into consideration in this paper. In appendix \ref{app-teh}, we demonstrate the membrane stress tensor \eqref{teh} remains correct when $a^{c}\neq0$.

\subsection{Fluid quantities}\label{subsect-quant}
Having set up the $2+1+1$ geometry of the static black bottle, the following steps are straightforward. In order to get the fluid quantities on the event horizon, we should calculate the stress tensor \eqref{teh} on the stretched horizon, and then decompose it to the form \eqref{tdec}.

By definition of the 3-dimensional extrinsic curvature \eqref{3k}, we find the nonzero components
\begin{equation}
K_{\tau\tau}=-\frac{1}{2}f^{1/2}\partial_{r}f,~~~~K_{\vartheta\vartheta}=\frac{2r}{1+\cos\vartheta}f^{1/2},~~~~K_{\varphi\varphi}=\frac{r\sin^2\vartheta}{2}(1+\cos\vartheta)f^{1/2}.
\end{equation}
One may check that
\begin{equation}
K_{ab}=\frac{1}{r}f^{1/2}\gamma_{ab}-\frac{1}{2}f^{-1/2}\left(\partial_{r}f\right)u_{a}u_{b}.
\end{equation}
The trace of the extrinsic curvature is
\begin{equation}
K=\frac{1}{2r}f^{-1/2}\left(4f+r\partial_{r}f\right).
\end{equation}
Substituting them into equation \eqref{teh}, we can get the stress tensor of the fluid on the stretched horizon. The nonvanishing components are
\begin{eqnarray}
\nonumber t_{\tau\tau}&=&-\frac{1}{4\pi Gr}f^{3/2},\\
\nonumber t_{\vartheta\vartheta}&=&\frac{r}{8\pi G(1+\cos\vartheta)}f^{-1/2}\left(2f+r\partial_{r}f\right),\\
t_{\varphi\varphi}&=&\frac{r\sin^2\vartheta}{32\pi G}(1+\cos\vartheta)f^{-1/2}\left(2f+r\partial_{r}f\right),
\end{eqnarray}
which are equivalent to
\begin{equation}
t_{ab}=\frac{1}{16\pi Gr}f^{-1/2}\left(2f+r\partial_{r}f\right)\gamma_{ab}-\frac{1}{4\pi Gr}f^{1/2}u_{a}u_{b}.
\end{equation}
Comparing it with equation \eqref{tdec} and inserting the renormalization parameter \eqref{alpha}, we arrive at some fluid quantities or their combinations
\begin{eqnarray}\label{fluqn2}
\nonumber&&\rho=-\frac{1}{4\pi Gr}f,~~~~\pi^{A}=0,~~~~\eta\sigma^{AB}=0,\\
&&p-\zeta\theta=\frac{1}{16\pi Gr}\left(2f+r\partial_{r}f\right).
\end{eqnarray}

Combining equations \eqref{fluqn1} and \eqref{fluqn2}, we can conclude that the fluid quantities on the event horizon are
\begin{eqnarray}\label{fluqn}
\nonumber&&\mathrm{Energy~density:~}\rho=-\frac{1}{4\pi Gr}f,\\
\nonumber&&\mathrm{Pressure:~}p=\frac{1}{16\pi G}\partial_{r}f,\\
\nonumber&&\mathrm{Momentum~density:~}\pi^{A}=0,\\
\nonumber&&\mathrm{Shear:~}\sigma^{AB}=0,\\
\nonumber&&\mathrm{Expansion:~}\theta=\frac{2}{r}f,\\
\nonumber&&\mathrm{Shear~viscosity:~}\eta=\frac{1}{16\pi G},\\
&&\mathrm{Bulk~viscosity:~}\zeta=-\frac{1}{16\pi G}.
\end{eqnarray}

The shear $\sigma^{AB}$ and the expansion $\theta$ can be calculated directly with the 2-dimensional extrinsic curvature \eqref{2k}, which turns out to be zero,
\begin{equation}
k_{AB}=0.
\end{equation}
According to equation \eqref{2kdec}, this means
\begin{equation}\label{sigth}
\sigma^{AB}=0,~~~~\theta=0.
\end{equation}
In the near-horizon limit $\alpha\rightarrow0$, it is consistent with equation \eqref{fluqn}.

\subsection{EF coordinates and surface gravity}\label{subsect-EF}
When building the membrane paradigm in details, the line element \eqref{metrictr} in spherical coordinates is very useful but not enough. In particular, it is unsuitable for computing the surface gravity $g_{\mathcal{H}}$ or studying the near-horizon limit of $n^{a}$. These difficulties can be overcome by introducing the Eddington-Finkelstein (EF) coordinates.

The ingoing EF coordinates $(v,r,\vartheta,\varphi)$ are related to the standard spherical coordinates $(\tau,r,\vartheta,\varphi)$ via
\begin{equation}
dv=d\tau+f^{-1}\left(dr-\frac{r^2\sin\vartheta}{L}d\vartheta\right)
\end{equation}
for the static black bottle solution \eqref{metrictr}. In the EF coordinates, the line element of a static black bottle takes the form
\begin{equation}\label{metricEF}
ds^2=-fdv^2+2dv\left(dr-\frac{r^2\sin\vartheta}{L}d\vartheta\right)+\frac{2r^2}{1+\cos\vartheta}d\vartheta^2+\frac{r^2\sin^2\vartheta}{2}(1+\cos\vartheta)d\varphi^2.
\end{equation}

Rewritten in the EF coordinates, the vector \eqref{l} becomes
\begin{equation}\label{lEF}
l^{a}\partial_{a}=\partial_{v}.
\end{equation}
It is not hard to check that $l^{a}$ is a Killing vector and is null at the event horizon. To make sure that it is the generator of event horizon, we should also check if it is normal to the event horizon. This can be done in the EF coordinates as follows. From equation \eqref{th}, we can see the normal vector of event horizon has the dual form proportional to
\begin{equation}
dr-\frac{r^2\sin\vartheta}{L}d\vartheta.
\end{equation}
At the same time, the dual form of vector \eqref{lEF} is
\begin{equation}
l_{a}dx^{a}=-fdv+dr-\frac{r^2\sin\vartheta}{L}d\vartheta.
\end{equation}
Obviously, on the event horizon $f=0$, this vector is the normal vector. The surface gravity at the event horizon can be safely derived with formula \eqref{surfg}. In the EF coordinates \eqref{metricEF}, it gives a definite answer
\begin{equation}\label{gBB}
g_{\mathcal{H}}=\frac{1}{2}\partial_{r}f.
\end{equation}
This answer is exactly what we have expected in equations \eqref{fluqn1}, \eqref{fluqn}.

For consistency, we should scrutinize equations \eqref{nulim} in the near-horizon limit $\alpha\rightarrow0$, taking $\alpha=f^{1/2}$. Translated into EF coordinates,
\begin{equation}
\alpha u^{a}\partial_{a}=\partial_{v},~~~~\alpha n^{a}\partial_{a}=\partial_{v}+f\partial_{r}.
\end{equation}
Both of them tend to vector \eqref{lEF} in the near-horizon limit satisfactorily.

\subsection{Fluid equations}\label{subsect-flueq}
In principle, the continuity equation \eqref{Ray} and the Navier-Stokes equation \eqref{NS} can be derived from the Gauss-Codazzi equations and the Einstein equation, by considering the fact that $n^{a}\nabla_{a}n_{b}\neq0$. With all fluid quantities in hand, here we will take a shortcut. We will insert the fluid quantities into equations \eqref{Ray}, \eqref{NS} to check them directly.

In the membrane paradigm with a cosmological constant \cite{Wang:2014aty}, we relegate the cosmological term to the bulk stress tensor
\begin{eqnarray}
\nonumber T_{ab}&=&-\frac{\Lambda}{8\pi G}g_{ab}\\
&=&\frac{3(1-b)}{16\pi GL^2}g_{ab}.
\end{eqnarray}
Projected to the null-null and null-transverse directions respectively, it becomes
\begin{equation}\label{flueq0}
T_{ab}l^{a}l^{b}=-\frac{3(1-b)}{16\pi GL^2}f,~~~~T_{ab}l^{b}\gamma^{a}_{~A}=0.
\end{equation}
For the rest terms in equations \eqref{Ray}, \eqref{NS}, starting from the fluid quantities in equation \eqref{fluqn}, lengthy but straightforward computations lead to the result
\begin{eqnarray}
\mathcal{L}_{l}\rho+\theta\rho+p\theta-2\eta\sigma_{AB}\sigma^{AB}-\zeta\theta^{2}&=&-\frac{1}{8\pi Gr}f\left(\frac{2}{r}f-\partial_{r}f\right),\label{flueq1}\\
\gamma_{A}^{~e}\mathcal{L}_{l}\pi_{e}+\pi_{A}\theta+p_{||A}-2(\eta\sigma^{B}_{~A})_{||B}-(\zeta\theta)_{||A}&=&\frac{\sin\vartheta}{4\pi GL}f(d\vartheta)_{A}.\label{flueq2}
\end{eqnarray}
We should warn that the above result cannot be gained without the help of equation \eqref{onf}. In view of equations \eqref{flueq0}, \eqref{flueq1}, \eqref{flueq2}, it is clear that the continuity equation \eqref{Ray} and the Navier-Stokes equation \eqref{NS} are well satisfied in the near-horizon limit $\alpha\rightarrow0$.

\section{Translated to Pleba\'nski-Demia\'nski-like coordinates}\label{sect-PD}
In the previous section, we have constructed the membrane paradigm of static black bottles in the spherical coordinates \eqref{metrictr}. All of the results can be accurately translated to the Pleba\'nski-Demia\'nski-like coordinates \eqref{metricxy} via relation \eqref{cotrans}. For future reference, let us accomplish the translation of equations \eqref{nugamma}, \eqref{nu}, \eqref{l},\eqref{alpha}, \eqref{fluqn}, \eqref{gBB}. These equations specify the geometric structure and the fluid quantities of the membrane.

Firstly, in the Pleba\'nski-Demia\'nski-like coordinates \eqref{metricxy}, we can rewrite equations \eqref{nugamma}, \eqref{nu}, \eqref{l} as
\begin{eqnarray}\label{nugammanul}
\nonumber&&n_{a}dx^{a}=\frac{\sqrt{2}L}{x-y}[-Q(y)]^{-1/2}dy,~~~~u_{a}dx^{a}=-\frac{\sqrt{2}L}{x-y}[-Q(y)]^{1/2}dt,\\
\nonumber&&\gamma_{AB}dx^{A}dx^{B}=\frac{2L^2}{(x-y)^2}\left[\frac{dx^2}{P(x)}+P(x)d\phi^2\right],\\
&&n^{a}\partial_{a}=\frac{x-y}{\sqrt{2}L}[-Q(y)]^{1/2}\partial_{y},~~~~u^{a}\partial_{a}=\frac{x-y}{\sqrt{2}L}[-Q(y)]^{-1/2}\partial_{t},~~~~l^{a}\partial_{a}=\frac{1}{2L}\partial_{t}.
\end{eqnarray}
Secondly,  in the Pleba\'nski-Demia\'nski-like coordinates, equations \eqref{alpha}, \eqref{fluqn}, \eqref{gBB} take the form
\begin{eqnarray}\label{alphafluqngBB}
\nonumber&&\alpha=\frac{1}{\sqrt{2}(x-y)}[-Q(y)]^{1/2},~~~~\rho=\frac{1}{8\pi GL(x-y)}Q(y),\\
\nonumber&&p=-\frac{1}{32\pi GL}\left[\frac{2}{x-y}Q(y)+\partial_{y}Q(y)\right],\\
\nonumber&&\pi^{A}=0,~~~~\sigma^{AB}=0,~~~~\theta=-\frac{1}{L(x-y)}Q(y),\\
&&\eta=\frac{1}{16\pi G},~~~~\zeta=-\frac{1}{16\pi G},~~~~g_{\mathcal{H}}=-\frac{1}{4L}\left[\frac{2}{x-y}Q(y)+\partial_{y}Q(y)\right].
\end{eqnarray}
What is more, the job in subsection \ref{subsect-EF} can be done by introducing the Pleba\'nski-Demia\'nski-like counterpart of EF coordinates. This should be straightforward and thus we leave it to the readers.

\section{Conclusion}\label{sect-con}
We have shown that the membrane paradigm of black holes can be applied to the static black bottle successfully and explicitly, although the normal vector of the stretched horizon violates the zero acceleration assumption in the literature. The success relies on two ``miracles'': first, the membrane stress tensor is not affected by the acceleration; and second, the fluid equations continue to hold. We have demonstrated the first miracle very generally in appendix \ref{app-teh}, and presented the second miracle for the static black bottle specifically in subsection \ref{subsect-flueq}.

In the membrane paradigm, the stretched horizon is interpreted as a fluid with certain dissipative properties, while the fluid quantities meet the continuity equation \eqref{Ray} and the Navier-Stokes equation \eqref{NS} in the near true horizon limit. That is to say, the membrane paradigm has established a relation between the black hole horizons and the hydrodynamics. Unfortunately, hitherto the relation has been confirmed only for black branes in asymptotically AdS spacetime. For black holes in asymptotically AdS spacetime, this is difficult because their horizons lack translational invariance. The black bottles suffer the same difficulty. However, if in the far future one discovers the dual field theory of black bottles in asymptotically AdS spacetime, its hydrodynamics limit should reproduce some of the results in this paper.

\begin{acknowledgments}
This work is supported by the National Natural Science Foundation of China (Grant No. 91536218), and in part by the Science and Technology Commission of Shanghai Municipality (Grant No. 11DZ2260700). T. W. is indebted to Shi-Ying Cai for encouragement and support.
\end{acknowledgments}

\appendix

\section{On the robustness of \eqref{teh}}\label{app-teh}
The assumption $a^{b}=0$ has been made in reference \cite{Parikh:1997ma} and other works. In this appendix, we will demonstrate that the membrane stress tensor \eqref{teh} is not modified if we get rid of this assumption. After examining the proof of equation (3.24) in reference \cite{Parikh:1997ma}, i.e. equation \eqref{teh} in our main text, we find the acceleration $a^{b}$ of normal vector appears exclusively in equation (A1) of reference \cite{Parikh:1997ma}. Therefore, to get the corrections of $a^{b}$ to stress tensor \eqref{teh}, we have to only revise equation (A1) of reference \cite{Parikh:1997ma}.

In equation (A1) of reference \cite{Parikh:1997ma}, the terms involving $a^{b}$ are
\begin{equation}
\int d^3x\sqrt{-h}\left[\left(n^{c}a^{b}+n^{b}a^{c}\right)\delta h_{bc}-h^{bc}n^{a}\delta h_{ab}a_{c}-a^{b}n^{a}\delta h_{ab}\right].
\end{equation}
Using the symmetry $\delta h_{ab}=\delta h_{ba}$, we can put it in the form
\begin{eqnarray}
\nonumber&&\int d^3x\sqrt{-h}\left(n^{a}a^{b}\delta h_{ac}-h^{bc}n^{a}\delta h_{ab}a_{c}\right)\\
\nonumber&=&\int d^3x\sqrt{-h}\left(g^{bc}-h^{bc}\right)n^{a}a_{c}\delta h_{ab}\\
&=&\int d^3x\sqrt{-h}n^{b}n^{c}n^{a}a_{c}\delta h_{ab}.
\end{eqnarray}
By the identify $n^{c}\nabla_{a}n_{c}=0$, it is easy to see $n^{c}a_{c}=0$ and thus the last line vanishes. Consequently, in the general case $a^{b}\neq0$, the terms involving $a^{b}$ does not contribute to the membrane stress tensor \eqref{teh}.

\end{document}